\documentclass[journal=jacsat,manuscript=article]{achemso}

\usepackage[version=3]{mhchem}
\usepackage{siunitx}
\usepackage{booktabs}
\usepackage[hidelinks]{hyperref}
\usepackage{adjustbox}

\DeclareSIUnit[quantity-product = ]\percent{\char`\%}
\DeclareSIUnit\angstrom{\text {Å}}
\DeclareSIUnit\electronmass{\text {\ensuremath {m}}_{0}}
\author{Ruiqi Wu}
\altaffiliation{These authors contributed equally to this work}
\affiliation{Department of Chemistry, Molecular Sciences Research Hub, White City Campus, Imperial College London, Wood Lane, London W12 0BZ, UK}

\author{JJ Acton}
\altaffiliation{These authors contributed equally to this work}
\affiliation{Department of Chemistry, Molecular Sciences Research Hub, White City Campus, Imperial College London, Wood Lane, London W12 0BZ, UK}

\author{Shirui Wang}
\affiliation{Department of Chemistry, Molecular Sciences Research Hub, White City Campus, Imperial College London, Wood Lane, London W12 0BZ, UK}

\author{Alex M. Ganose}
\affiliation{Department of Chemistry, Molecular Sciences Research Hub, White City Campus, Imperial College London, Wood Lane, London W12 0BZ, UK}
\email{a.ganose@imperial.ac.uk}

\title{On the possibility of hybrid chalcogenide perovskite photovoltaics}

\begin{document}

\begin{abstract}
Chalcogenide perovskites are an emerging class of photovoltaic absorbers offering stable, lead-free structures and promising optoelectronic properties.
To date, the literature on chalcogenide perovskites has focused primarily on fully inorganic systems such as \ce{BaZrS3}.
This contrasts with the halide perovskites, for which hybrid organic-inorganic systems exhibit record performance.
In this work, we assess the viability of hybrid chalcogenide perovskite absorbers using first-principles calculations.
We screen a wide range of monovalent and divalent organic cations within the A-site to evaluate their electronic, optical, and thermodynamic properties.
Our analysis reveals that the majority of candidates are structurally unstable; however, we identify the hydrazinium cation (\ce{N2H6^{2+}}) as a unique candidate that maintains a stable perovskite structure.
Specifically, we identify \ce{N2H6ZrSe3} as the most promising candidate, exhibiting a quasi-direct band gap of \SI{1.31}{eV} and a theoretical maximum efficiency of \SI{24.5}{\percent} for a \SI{200}{\nm} thin film.
This study represents the first comprehensive computational report on hybrid chalcogenide perovskites, opening new avenues for the development of Earth-abundant photovoltaic materials.
\end{abstract}

\section{Introduction}

As solar energy deployment accelerates, the development of highly efficient, non-toxic, and Earth-abundant photovoltaic materials is paramount.\cite{pv_roadmap}
While silicon dominates the current market, alternative thin-film technologies are often constrained by a reliance on toxic (e.g., Cd, Pb) or scarce (e.g., Te, In) elements, hindering their scalability.\cite{Haegel2019,Li2020LeadSequestration}
Over the last decade, perovskites have gained significant attention as a promising solution for next-generation photovoltaics.
Their unique chemical and physical properties stem from the \ce{ABX3} crystal structure, characterised by a three-dimensional network of corner-sharing \ce{BX6} octahedra with an A-site cation occupying the cuboctahedral cavity.

The most widely researched perovskite photovoltaic devices are lead halide hybrid organic-inorganic perovskites, which comprise a monovalent organic A-site cation, divalent Pb B-site cation, and a halide X-site anion.\cite{frost_atomistic_2014}
They are generally facile to synthesise using low-cost, solution-processable methods and are chemically tuneable, enabling high photovoltaic performance.\cite{giovanni_optical-spin_2017,jeon_compositional_2015}
To date, single-junction perovskite cells have reached record efficiencies of \SI{26.7}{\percent}.\cite{noauthor_interactive_nodate,maalouf_comprehensive_2023}
Methylammonium lead iodide (\ce{CH3NH3PbI3}, or \ce{MAPbI3}) is the most extensively studied perovskite absorber due to its exceptional optoelectronic properties.
It displays a direct band gap of approximately \SI{1.55}{eV}, low exciton binding energies, high defect tolerance, and remarkable charge carrier diffusion lengths.\cite{baikie_synthesis_2013,ganose_beyond_2016,bati_next-generation_2023,dong_electron-hole_2015}
However, while hybrid lead halide perovskites continue to break efficiency records, their poor environmental stability and reliance on toxic lead have prevented wide-scale commercialisation.\cite{yin_halide_2015}

To address these challenges, chalcogenide materials, comprising group 16 anions (\ce{S^{2-}}, \ce{Se^{2-}}), have been integrated into the \ce{ABX3} perovskite architecture.
This approach aims to combine the stability provided by strongly covalent metal-chalcogenide bonds with the favourable electronic dimensionality of the perovskite lattice.\cite{sopiha_chalcogenide_2022}
\ce{BaZrS3} is the most extensively studied chalcogenide perovskite.
It crystallises in the orthorhombic \ce{GdFeO3}-type (\textit{Pnma}) structure and exhibits exceptional thermodynamic and environmental stability.\cite{nishigaki_extraordinary_2020}
The corner-sharing \ce{ZrS6} octahedra provide large band dispersion, leading to low effective masses and high charge carrier mobilities.
Furthermore, \ce{BaZrS3} displays a high absorption coefficient ($>$\SI{e5}{\per\cm}) near the absorption onset and a direct band gap of $\approx$\SI{1.78}{eV}.\cite{nishigaki_extraordinary_2020}
It is lead-free and several synthesis routes exist to form polycrystalline thin films.\cite{thakur_recent_2023,lelieveld_sulphides_1980}
Crucially, chalcogenides such as \ce{BaZrS3} exhibit superior stability in the presence of moisture and elevated temperatures compared to their halide counterparts.
Additionally, they benefit from reduced toxicity, a primary advantage over lead halide and cadmium telluride based photovoltaics.

To date, all experimental reports have focused on fully inorganic compounds.
However, analogous to the halides, substituting the inorganic A-site with an organic molecule offers a route to tune optoelectronic properties, increase structural degrees of freedom, and potentially reduce synthesis costs.
Currently, reports on hybrid chalcogenide perovskites remain scarce.
\citet{sun_chalcogenide_2015} theoretically investigated the substitution of the inorganic A-site with the organic hydrazinium ion (\ce{N2H6^{2+}}) to form \ce{[N2H6]ZrS3}, predicting a direct band gap of \SI{1.68}{eV} and a theoretical efficiency of \SI{29}{\percent} using the Heyd-Scuseria-Ernzerhof (HSE06) hybrid functional.
\textit{Ab initio} molecular dynamics at \SI{423}{K} further suggested the structure is dynamically stable.

In this work, we systematically assess the viability of hybrid organic-inorganic chalcogenide perovskites for photovoltaic applications using first-principles calculations.
We screen a wide range of charge-balanced organic cations to identify those that yield geometrically stable perovskite structures.
The identified candidates are relaxed to their ground state and their optoelectronic properties evaluated using hybrid density functional theory.
We identify the HZ(Zr,Hf)(S,Se)$_3$ family (where HZ = hydrazinium) as a unique candidate class, exhibiting band gaps within the ideal range for solar absorption.
Subsequently, we assess the thermodynamic and dynamical stability of the most promising candidates using convex hull analysis and molecular dynamics.
Finally, band alignment calculations are performed to identify suitable contact materials for device integration.

\section{Computational Methodology}

All calculations were performed using the Kohn-Sham DFT framework within the Vienna Ab initio Simulation Package (VASP), employing the projector augmented wave method.\cite{kresse_efficiency_1996, blochl_projector_1994}
Structural optimisations utilised the generalised gradient approximation (GGA) of Perdew, Burke, and Ernzerhof,\cite{perdew_generalized_1996} specifically the revised version adapted for solids (PBEsol).\cite{perdew_restoring_2008}
Atoms were relaxed until Hellmann-Feynman forces were below \SI{0.01}{eV \angstrom^{-1}} and the total energy was converged to \SI{1e-5}{\eV}.
The plane-wave energy cut-off and $k$-point mesh were converged to a tolerance of \SI{5}{meV/atom} and \SI{1}{meV/atom} respectively, yielding a converged cut-off of \SI{500}{\eV} with the converged $k$-point meshes provided in Supplementary Table S5.
During geometry optimisations, the cut-off was increased to \SI{650}{\eV} to account for Pulay stress errors.\cite{pulay_ab_1969}
The HSE06 functional was used to calculate the band structures, density of states (DOS) and optical absorption, which were plotted using the \textsc{sumo} package.\cite{heyd_hybrid_2003, m_ganose_sumo_2018}
Use of a hybrid functional is essential to correct for band gap underestimation seen in semilocal GGA functionals, as previously demonstrated across many emerging photovoltaics.\cite{wu_phosphide_2023,wu_relativistic_2023}
To further provide evidence of our choice of exchange correlation functionals, we benchmark their performance on the geometric and optoelectronic properties of \ce{BaZrS3} and \ce{BaHfS3} in Section S1 of the Supplementary Information.

Structural stability and phase formation in perovskites are described by the Goldschmidt tolerance factor and octahedral factor, which quantify structural distortion and octahedral stability, respectively.
The Goldschmidt tolerance factor ($t$)\cite{goldschmidt_gesetze_1926} is defined as
\begin{equation}
    t=\dfrac{R_A+R_X}{\sqrt{2}(R_B+R_X)},
\end{equation}
where $R_A, R_B$ and $R_X$ are the ionic radii for the respective sites. A tolerance factor of 1 indicates preference for a cubic structure, while $0.8 < t < 1$ is often considered a stable range for perovskites.
The octahedral factor ($\mu$) is a measure of whether the B-site cation is large enough to prevent the octahedron from collapsing and is satisfied when $0.4\leq\mu\leq0.8$.
It is defined as
\begin{equation}
    \mu=\frac{R_B}{R_X}.
\end{equation}
We derived the effective organic A-site cation radii, $r_A$, following the approach of \citet{cheetham_organic_radii_2014} Assuming unrestricted rotation about the centre of mass, cations are treated as rigid spheres with
\begin{equation}
    r_A = r_{\text{mass}} + r_{\text{ion}},
    \label{eq:radii}
\end{equation}
where $r_{\text{mass}}$ represents the distance from the centre of mass to the most distant non-hydrogen atom and $r_{\text{ion}}$ denotes the ionic radius of that specific atom.

To generate hybrid organic-inorganic chalcogenide perovskites, molecular substitution was implemented using the \textsc{pymatgen}\cite{ong_python_2013} package.
Organic cations with oxidation states of $+1$ or $+2$ were used to replace the inorganic A-site atom, ensuring charge neutrality.
Thermodynamic stability was evaluated using the \textsc{doped}\cite{kavanagh_doped_2024} software package, employing the convex hull formalism to assess stability relative to competing phases. Competing phases were identified by querying the Materials Project\cite{jain_commentary_2013} within the host chemical space and selecting phases that lie on the convex hull. For each target compound and its competing phases, the ground state energies were calculated using a converged $k$-mesh and the PBEsol functional. The thermodynamic stability was determined by computing the energy above hull of the target phase with respect to the competing phases, where an energy above the hull of \SI{0}{\eV} indicates stability, and a positive value implies a driving force to decompose into lower energy phases. 

\textit{Ab initio} molecular dynamics (AIMD) simulations were performed on $2\times 4\times 4$ supercells of the orthorhombic structure to assess stability at finite temperature.
Molecular dynamics simulations were performed in the isothermal–isobaric (NPT) ensemble using a Langevin thermostat with a Parrinello–Rahman barostat at \SI{300}{K}, employing a timestep of \SI{0.5}{\fs} and a total simulation time of \SI{10}{\ps}.
A single $k$-point at $\Gamma$ was used for all AIMD simulations.
To further reduce computational resources during AIMD simulations, we used on-the-fly machine-learning force fields (MLFF) employing Gaussian approximation potentials as implemented in VASP.\cite{jinnouchi_phase_2019,jinnouchi_--fly_2019}
The potentials were configured following the VASP 6.5.0 default settings.
To identify the ground state cation orientation, we performed an annealing strategy composed of an initial temperature ramp to \SI{450}{\kelvin}, followed by an equilibration run at \SI{450}{\kelvin} for \SI{10}{\ps}, and subsequent annealing to \SI{10}{\kelvin}.
All three stages were performed for \SI{10}{\ps} each, with a \SI{0.5}{\fs} timestep.
The final structure at the end of the annealing run was relaxed using the trained on-the-fly forcefield, symmetrised using \textsc{spglib}\cite{togo_spglib_2024}, and relaxed again using PBEsol to obtain the final ground-state structure.
Tilting mode analysis was performed from molecular dynamics trajectories using the \textsc{PDynA} package.\cite{xia_structural_2023}

The high-frequency dielectric response ($\varepsilon_{\infty}$) was calculated using HSE06 through the frequency-dependent microscopic polarisability matrix as implemented in VASP.\cite{gajdos_linear_2006}
The low-frequency ionic dielectric response ($\varepsilon_{\text{ionic}}$), which accounts for the response of phonons under an applied electric field, was calculated using density functional perturbation theory with the PBEsol functional.
The static dielectric constant is given by $\varepsilon_0= \varepsilon_{\infty}+\varepsilon_{\text{ionic}}$.
The conductivity effective mass, taking into account a Fermi-Dirac weighted average of all bands near the band edge, was obtained using the \textsc{AMSET} package.\cite{ganose_efficient_2021}

To evaluate the upper radiative limit of thickness-dependent energy conversion, the Spectroscopic Limited Maximum Efficiency (SLME) detailed balance metric was employed, based on the optical properties calculated using HSE06.\cite{zunger_slme_2012}
Band alignment calculations employed a slab structure generated using the \textsc{surfaxe} package\cite{brlec_surfaxe_2021} with slab and vacuum thicknesses of \SI{30}{\angstrom}.
The electron affinity and ionisation potentials were determined based on alignment between the core and vacuum levels of the slab, and core and valence band maximum of the bulk.
All band alignment calculations were performed using the HSE06 functional.

\section{Results and Discussion}

\subsection{Materials Screening Strategy}

\begin{figure}[t]
    \centering
    \includegraphics[width=1\textwidth]{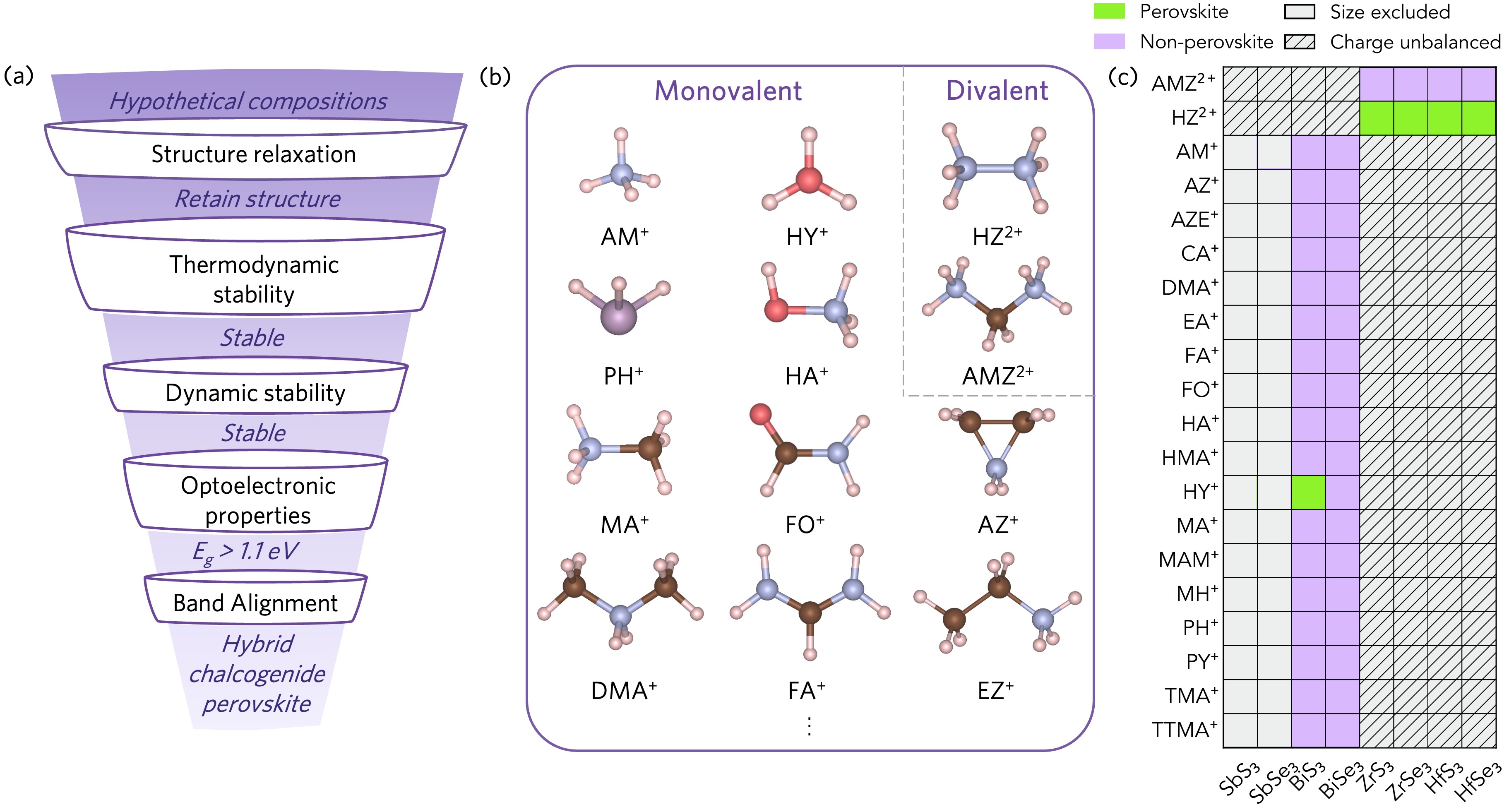}
    \caption{Overview of the material screening procedure used to identify the hybrid chalcogenide candidates from the 50 initial compounds. (a) Detailed screening workflow based on stability and optoelectronic properties. (b) The monovalent and divalent organic molecules studied in this work. (c) Structural stability after geometry optimisation using the PBEsol functional, where green and purple blocks refer to the perovskite and non-perovskite structure, respectively.}
    \label{fig:workflow}
\end{figure}

To identify viable hybrid chalcogenide perovskites, we employed a hierarchical screening workflow, progressively filtering candidates based on chemical constraints, structural integrity, and thermodynamic stability.
\textit{Ab initio} molecular dynamics were employed to assess dynamical stability and capture cation-driven octahedral tilting.
For stable compounds, we evaluate their optoelectronic properties relevant to photovoltaic operation, including band gaps, carrier effective masses, strength of optical absorption, and predicted maximum theoretical photovoltaic performance.
Finally, band alignment analysis is performed to inform the selection of suitable contact layer materials for electron and hole extraction.
The overall process is summarised in Figure \ref{fig:workflow}a.

Our chemical search space was defined by the \ce{ABX3} perovskite stoichiometry.
For the B-site, we identified two cation groups: trivalent pnictogens (\ce{Sb}, \ce{Bi}) and tetravalent transition metals (\ce{Zr}, \ce{Hf}).
These elements were prioritised due to their relatively high Earth abundance, reduced toxicity compared to Pb or Cd, and their use in emerging chalcogenide absorbers such as \ce{Sb2Se3} and \ce{BaZrS3}.\cite{wang_lone_2022a}
The choice of B-site dictates the charge requirements for the organic A-site to maintain neutrality.
The \ce{B^{3+}} series (\ce{Sb}, \ce{Bi}) requires a monovalent (\ce{A^{+}}) cation, while the \ce{B^{4+}} series (\ce{Zr}, \ce{Hf}) requires a divalent (\ce{A^{2+}}) cation.

To begin, a stable ground-state bulk structure is required as a reference framework into which the A-site cation can be substituted.
Given that perfect cubic structures (e.g., \ce{SrTiO3}) are rare in later-group chalcogenide perovskites, we used the orthorhombic \ce{BaZrS3} ground-state phase (\textit{Pnma}) as a reference structure.\cite{sopiha_chalcogenide_2022}
The initial structure was obtained from the Materials Project database.\cite{jain_commentary_2013}
Candidate organic cations were identified by querying the PubChem database\cite{kim_pubchem_2023} for molecules satisfying the charge neutrality requirements imposed by the B-site monovalent cations for Sb/Bi and divalent cations for Zr/Hf.
We applied a molar mass cut-off of \SI{75}{\gram\per\mol} as a proxy for ionic size to maximise the probability of fitting within the steric constraints of the chalcogenide cage.
The full set of organic cations considered is provided in Table~\ref{tab:organic_cations}.

Sourcing suitable divalent organic cations presented a significant chemical challenge, since they are often intrinsically unstable due to electron deficiencies and high ionisation energies.
Consequently, only two realistic divalent candidates were identified: hydrazinium \ce{(NH3-NH3)^{2+}} and azaniumylmethylazanium \ce{(NH3-CH2-NH3)^{2+}}.
Due to this scarcity, monovalent cations dominate our list as they are far more abundant (see Table \ref{tab:organic_cations}).
In total, we identified 84 unique hypothetical compositions for evaluation.
As illustrated in the matrix in Figure \ref{fig:workflow}(c), the combination of these species defines the screening phase space.
The hatched grey blocks correspond to charge-imbalanced compositions.
The remaining grey blocks represent size-excluded compositions which we define as compounds with an octahedral factor less than 0.34. This resulted in all of the Sb compounds being screened out, with a final set of 46 compositions selected for further study.
The full list of tolerance and octahedral factors for all 84 compounds is provided in Table S4 of the Supplementary Information.

\begin{table}[t]
\centering
\caption{Organic cations used as A-site molecules in the \ce{ABX3} perovskite structure, including the molar mass ($M$) in g/mol. The cation radius was calculated according to Eqn.~\ref{eq:radii}}
\begin{adjustbox}{width=\textwidth}
\begin{tabular}{l l l c c}
\toprule
Cation & Formula & Label & Molar Mass (g/mol) & $R_A$ (pm)\\ \midrule
Ammonium & \ce{[NH4]+} & AM\textsuperscript{+} & 18.04 & 146 \\
Hydronium & \ce{[H3O]+} & HY\textsuperscript{+} & 19.02 & 220\\
Methaniminium & \ce{[CH2NH2]+} & MAM\textsuperscript{+} & 30.05 & 216 \\
Methylammonium & \ce{[CH3NH3]+} & MA\textsuperscript{+} & 32.06 & 217\\
Hydrazinium & \ce{[NH3NH2]+} & HZ\textsuperscript{+} & 33.05 & 220 \\
Hydroxylammonium & \ce{[NH3OH]+} & HA\textsuperscript{+} & 34.04 & 216 \\
Phosphonium & \ce{[PH4]+} & PH\textsuperscript{+} & 35.00 & 212 \\
Aziridinium & \ce{[NH2(CH2)2]+} & AZ\textsuperscript{+} & 44.08 & 238\\
Formamidinium & \ce{[CH(NH2)2]+} & FA\textsuperscript{+} & 45.08 & 253\\
Ethylammonium & \ce{[CH3CH2NH3]+} & EA\textsuperscript{+} & 45.10 & 274\\
Formamide & \ce{[NH3COH]+} & FO\textsuperscript{+} & 46.06 & 258\\
Dimethylazanium & \ce{[NH2(CH3)2]+} & DMA\textsuperscript{+} & 46.09 & 272\\
Methylhydrazinium & \ce{[CH3NH2NH3]+} & MH\textsuperscript{+} & 47.08 & 275\\
Azetidinium & \ce{[NH2(CH2)3]+} & AZE\textsuperscript{+} & 58.10 & 250\\
Hydroxy(methylidene)azanium & \ce{[CH2NHOH]+} & HMA\textsuperscript{+} & 58.06 & 275 \\
Cyclopropylammonium & \ce{[NH3CH(CH2)2]+} & CA\textsuperscript{+} & 58.10 & 293\\
Trimethylazanium & \ce{[NH(CH3)3]+} & TMA\textsuperscript{+} & 60.12 & 294\\
Pyrrolidinium & \ce{[NH2(CH2)4]+} & PY\textsuperscript{+} & 72.13 & 278\\
Tetramethylazanium & \ce{[N(CH3)4]+} & TTMA\textsuperscript{+} & 74.14 & 292\\ \midrule
Hydrazinium & \ce{[NH3NH3]^{2+}} & HZ\textsuperscript{2+} & 34.06 & 217 \\
Azaniumylmethylazanium & \ce{[CH2(NH3)2]^{2+}} & AMZ\textsuperscript{2+} & 48.11 & 267\\ \bottomrule
\label{tab:organic_cations}
\end{tabular}
\end{adjustbox}
\end{table}

\subsection{Structural Stability}

The structural viability of the 46 hypothetical compounds was assessed via geometry optimisation using the PBEsol functional.
A candidate was considered structurally stable only if it retained the corner-sharing \ce{BX6} octahedral network characteristic of the perovskite phase upon relaxation.
As shown by the purple blocks in Figure \ref{fig:workflow}(c), the majority of compounds do not produce a stable perovskite structure, instead forming non-perovskite phases due to bond breaking resulting from large lattice distortions.

In the monovalent \ce{A^{+}Bi^{3+}X3} series, structural instability was pervasive.
As previously discussed, only Bi was considered as the B-site cation due to its larger ionic radius (\SI{0.76}{\angstrom}) compared to Sb (\SI{0.60}{\angstrom}) in the $5^+$ charge state, increasing the octahedral factor closer to $0.40$.\cite{shannon_effective_1969}
Among the monovalent A-site cations, only hydronium (\ce{HY+}) retained the corner-sharing perovskite framework.
This cation has a small effective radius ($R_A = \SI{140}{pm}$) and low molecular weight, which favours accommodation within the A-site cavity, consistent with results reported by \citet{park_learn-and-match_2019}.
In contrast, structural stability declines sharply for effective radii exceeding \SI{220}{pm}, with no other monovalent cations remaining stable in the perovskite structure.

In the divalent \ce{A^{2+}B^{4+}X3} series (B = \ce{Zr}, \ce{Hf}), we observed a strict size cut-off.
The azaniumylmethylazanium compounds (\ce{AMZ^{2+}}), containing a methylene bridge (\ce{-CH2-}) between the amine groups, decomposed into low-energy non-perovskite structures.
The only survivor was the hydrazinium cation (\ce{N2H6^{2+}}, HZ), which lacks the methylene spacer and possesses a smaller effective radius of \SI{217}{pm}.
This can be seen in Table \ref{tab:organic_tf}.
The HZ cation successfully retained the orthorhombic perovskite structure for both Zr and Hf-based compounds.

All surviving perovskite structures exhibited Goldschmidt tolerance factors in the range $0.90 < t < 1.11$, consistent with a distorted pseudocubic or orthorhombic coordination,\cite{goldschmidt_gesetze_1926} and an octahedral factor $\mu>0.36$.
Although an octahedral factor of $>0.40$ is desirable, compounds can remain stable with values slightly below this threshold, as highlighted by \ce{BaZrS3} ($\mu=0.39$).

\begin{table}[t]
\centering
\caption{Effective organic cation radius ($R_A$) and ionic radii ($R_B, R_X$) used to calculate the tolerance factor ($t$) and octahedral factor ($\mu$) as well as the average octahedral tilting angle for the chalcogenide perovskites.}
\begin{adjustbox}{width=0.8\textwidth}
\begin{tabular}{lccccccc}
\toprule
Compound & $R_A$ (pm) & $R_B$ (pm) & $R_X$ (pm) & $t$ & $\mu$ & Tilt angle (\si{\degree}) \\
\midrule
\ce{HYBiS3}   & 140 & 76 & 184 & 0.90 & 0.41 & 3.08\\
\ce{HZZrS3}   & 217 & 72 & 184 & 1.11 & 0.39 & 9.72\\
\ce{HZHfS3}   & 217 & 71 & 184 & 1.11 & 0.39 & 9.11\\
\ce{HZZrSe3}  & 217 & 72 & 198 & 1.09 & 0.36 & 11.86\\
\ce{HZHfSe3}  & 217 & 71 & 198 & 1.09 & 0.36 & 11.30\\
\bottomrule
\label{tab:organic_tf}
\end{tabular}
\end{adjustbox}
\end{table}

\subsection{Thermodynamic Stability}

Structural preservation after relaxation is only a weak proxy for stability.
The material must also be thermodynamically stable against decomposition into competing phases.
We evaluated the thermodynamic stability of the structurally surviving candidates by calculating their energy above the convex hull ($E_\text{hull}$) using PBEsol.
Thermodynamic properties were analysed based on formation energies and stability against competing phases, where negative formation energies indicate stability with respect to elemental ground states.\cite{bartel_review_2022}

The monovalent \ce{HYSbS3} exhibited a large $E_\text{hull}$ value over \SI{400}{meV/atom}, indicating thermodynamic instability.
Despite retaining the perovskite geometry, this material is expected to decompose into lower-energy binary chalcogenides and volatiles under any realistic synthesis conditions.
Consequently, none of the monovalent candidates pass both the thermodynamic and structural stability checks, and the entire monovalent series was excluded from further consideration.

In contrast, the divalent HZ-based family (\ce{HZZrS3}, \ce{HZZrSe3}, \ce{HZHfS3}, \ce{HZHfSe3}) was found to lie on the convex hull ($E_\text{hull} = \SI{0}{eV/atom}$).
This confirms that these hybrid hydrazinium chalcogenides represent the thermodynamic ground states for their respective compositions, suggesting they could be synthetically accessible.
Based on this screening process, only the HZ family was selected for detailed optoelectronic characterisation.

\subsection{Dynamic Stability}

\begin{figure}[t]
    \centering
    \includegraphics[width=\textwidth]{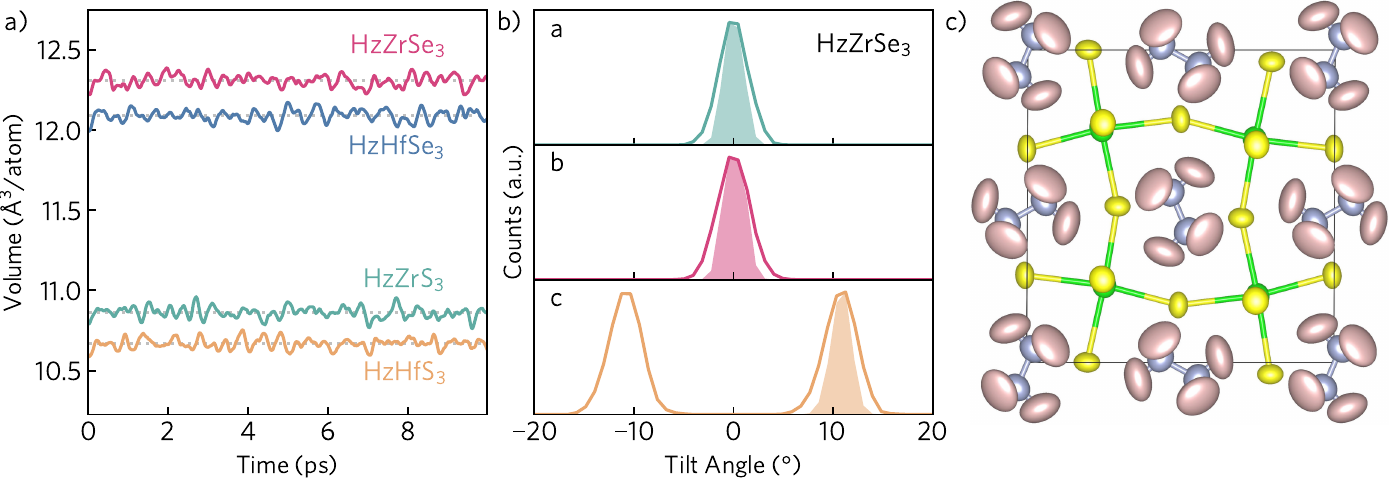}
    \caption{Dynamical stability and structural analysis of hydrazinium chalcogenide perovskites at \SI{300}{\kelvin}. (a) Time evolution of the volume per atom for \ce{HZZrS3}, \ce{HZZrSe3}, \ce{HZHfS3}, and \ce{HZHfSe3} over a \SI{10}{\ps} MLFF-MD trajectory, indicating structural stability. (b) Dynamic distribution of octahedral tilt angles for \ce{HZZrSe3} decomposed into the principal lattice vectors. Solid lines show the distribution of tilting, while the shaded region represents the corresponding nearest-neighbour tilts along the same direction. The shaded area in the positive angle range reveals an in-phase correlation along the $c$-axis ($a^0b^0c^+$ in Glazer notation). (c) Thermal ellipsoid plot of \ce{HZZrSe3} derived from the MD trajectory (80\% probability), shown in the a-b plane}
    \label{fig:aimd}
\end{figure}

To assess the robustness of the perovskite framework at finite temperatures, we performed MLFF-accelerated molecular dynamics simulations.
Figure \ref{fig:aimd}a presents the evolution of the cell volume for the \ce{HZZrS3}, \ce{HZZrSe3}, \ce{HZHfS3}, and \ce{HZHfSe3} compounds over a \SI{10}{\ps} trajectory at \SI{300}{\kelvin}.
In all cases, the volume fluctuates around a constant equilibrium value with no drift.
This indicates the absence of phase transitions or decomposition pathways, further suggesting that the hybrid hydrazinium chalcogenides are dynamically stable at room temperature.

To further understand the dynamics of octahedral framework and organic cations and gain insight into the local symmetry breaking, we extracted the instantaneous octahedral tilting angles ($\theta$) and their spatial correlations from the MD trajectories using the \textsc{PDynA} package.\cite{xia_structural_2023}
As shown in the histogram for \ce{HZZrSe3} (Figure \ref{fig:aimd}b), the tilting behaviour is highly anisotropic.
Distributions for the $a$ and $b$ axes are centred at zero, indicating no permanent static tilt along these directions.
In contrast, the $c$-axis exhibits a distribution with two peaks with a mean magnitude of \SI{11.38}{\degree} (averaged values for all compounds are detailed in Table~\ref{tab:organic_tf}).
Crucially, the spatial correlation function along the $c$-axis is positive, indicating an in-phase rotation of adjacent octahedra.
This tilting pattern corresponds to the $a^0b^0c^+$ Glazer system,\cite{glazer_tilt_1972} which lowers the symmetry of the aristotype structure to the orthorhombic \textit{Pbam} space group. In the fully inorganic analogue, the same $a^0b^0c^+$ tilt system is typically associated with the tetragonal \textit{P4/mbm} space group, while A-site molecular ordering and hydrogen bonding in hybrid chalcogenide introduce additional symmetry breaking, lowering the overall symmetry to \textit{Pbam}. This is in contrast to \ce{BaZrS3}, which crystallises in the \textit{Pnma} structure characterised by a complex three-axis tilting system ($a^-b^+a^-$).\cite{nishigaki_extraordinary_2020}
The reduction in tilting dimensionality in the hybrid systems suggests that the anisotropic shape and steric bulk of the hydrazinium cation effectively suppress rotations along the in-plane axes.
The full set of lattice parameters for the relaxed stable chalcogenide perovskites are provided in Supplementary Table S6.

The limited rotation of the hydrazinium cation is further elucidated by the thermal displacement analysis presented in Figure \ref{fig:aimd}c for \ce{HZZrSe3}.
The thermal ellipsoids for the Zr and Se sites are relatively small, consistent with a stiff inorganic framework with limited degrees of rotation.
For the organic A-site, the nitrogen atoms of the hydrazinium molecule also exhibit small thermal ellipsoids.
While the hydrogen atoms display larger displacement parameters, the anchoring of the N--N axis indicates that the molecule is fixed by a hydrogen-bonding network with the chalcogen anions, as seen at low temperatures in other hybrid perovskite compounds.
This further highlights that the \ce{N2H6^{2+}} molecule is not rotationally disordered but is instead locked into a specific orientation within the cuboctahedral cavity, in contrast to the methylammonium cation in halide perovskites (\ce{MAPbI3}), which display considerably more rotational degrees of freedom at room temperature.\cite{xia_phase_2025}

\subsection{Electronic Properties}

\begin{figure}[t]
    \centering
    \includegraphics[width=0.9\textwidth]{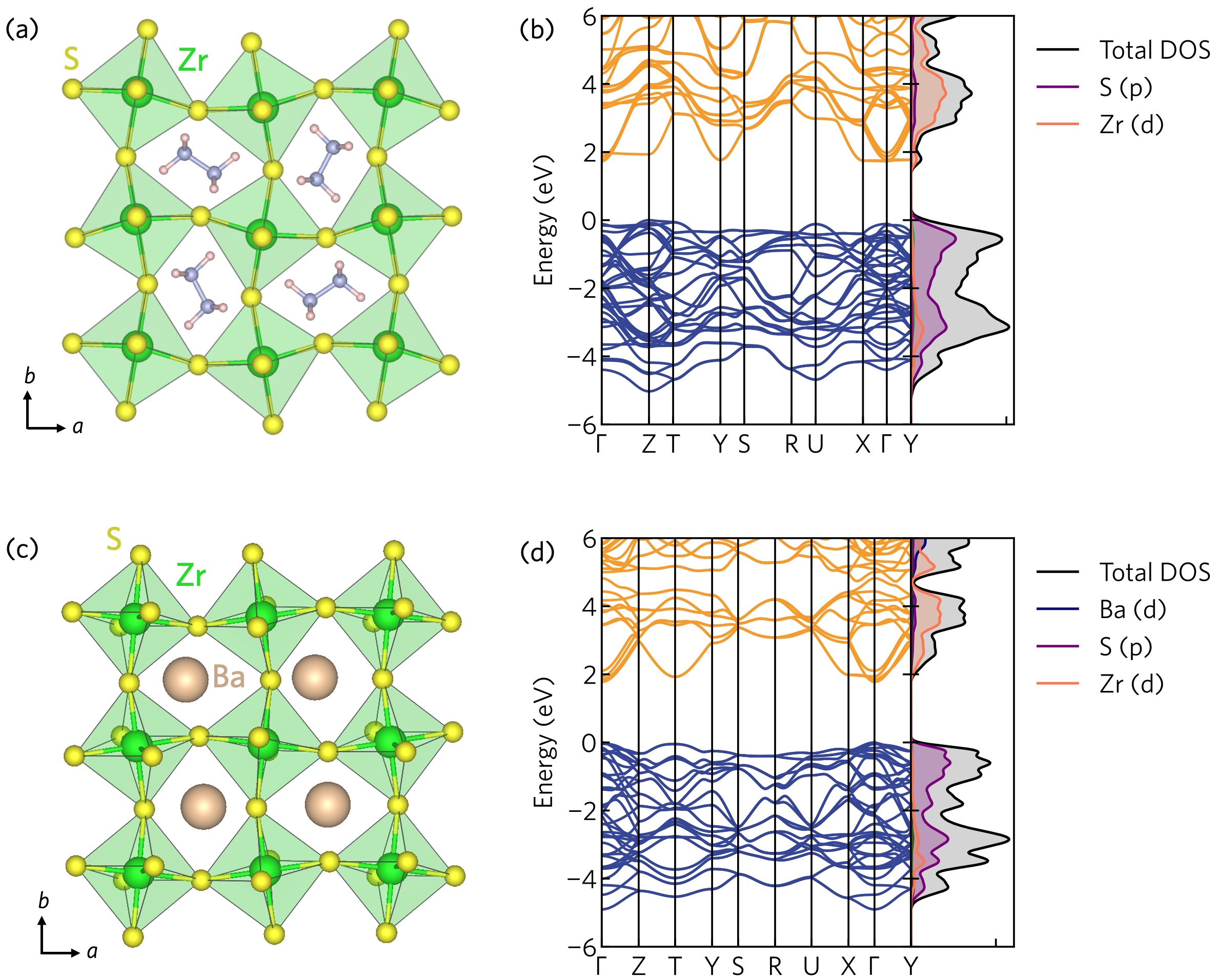}
    \caption{Comparison of crystal structure and electronic properties between hybrid and inorganic chalcogenide perovskites. (a) Crystal structure and (b) band structure of \ce{HZZrS3}. (c) Crystal structure and (d) band structure of  \ce{BaZrS3}.}
    \label{fig:band_structure}
\end{figure}

Having established the thermodynamic and dynamic stability of the hydrazinium-based perovskites, we now turn to their electronic structure to assess their viability as photovoltaic absorbers.
Figure \ref{fig:band_structure} presents the crystal structure and electronic band structure of \ce{HZZrS3} calculated using the HSE06 hybrid functional.
The corresponding data for the isoelectronic and isostructural analogues (\ce{HZZrSe3}, \ce{HZHfS3}, and \ce{HZHfSe3}) are provided in Supplementary Figure S5 and display similar features.

The calculated density of states (DOS) indicates that the optoelectronic behaviour of the \ce{HZBX3} series is governed almost exclusively by the inorganic framework.
The valence band (VB) edge is dominated by chalcogen $p$ states, while the conduction band (CB) is composed primarily of B-site transition metal $d$ orbitals.
Crucially, the N $s$ and $p$ orbitals of the hydrazinium cation make negligible contributions to the frontier states.
This confirms that the organic moiety does not participate directly in charge transport.
Instead, its role is structural, controlling the degree of lattice distortion and ensuring charge neutrality, with indirect impacts on electronic properties.
This behaviour is consistent across the series, although some Hf $p$ contributions are observed deep in the valence band (\SI{-4}{eV}) for the Hf-analogues, reflecting their semi-core nature.

A key consequence of the hybrid A-site substitution is a modification of the band edge momentum.
As illustrated in Figure \ref{fig:band_structure}, while the inorganic \ce{BaZrS3} is a direct gap semiconductor, the hybrid sulphide compounds (\ce{HZZrS3} and \ce{HZHfS3}) exhibit an indirect fundamental gap, with the valence band maximum (VBM) located at the Z-point and the conduction band minimum (CBM) at the $\Gamma$-point.
We attribute this transition to the specific symmetry-breaking distortions induced by the anisotropic \ce{HZ^{2+}} cation.
In the inorganic \ce{BaZrS3} structure (\textit{Pnma}), the octahedral tilting pattern preserves the VBM at $\Gamma$.
However, the hybrid \ce{HZZrS3} structure adopts a distinct tilting mode driven by the rigid, hydrogen-bonded hydrazinium network.
This alters the orbital overlap in the (001) plane, enhancing the antibonding S $3p$--S $3p$ interactions at the Z-point relative to the inorganic phase.
This interaction pushes the valence band energy at the Z-point above that at $\Gamma$, resulting in an indirect transition.
Crucially, the magnitude of this indirectness ($\Delta E_{\Gamma-Z}$) is small and composition-dependent (Fig.~\ref{fig:optics}a).
For \ce{HZZrS3}, the direct transition at $\Gamma$ is only \SI{0.12}{\eV} higher than the fundamental indirect gap.
Substituting S with Se significantly reduces this difference to just \SI{0.05}{\eV} in \ce{HZZrSe3}, rendering the band gap effectively direct at room temperature due to thermal broadening.
Conversely, the Hf-analogues exhibit slightly larger differences, with \ce{HZHfS3} and \ce{HZHfSe3} showing $\Delta E_{\Gamma-Z}$ values of \SI{0.15}{\eV} and \SI{0.11}{\eV}, respectively.

Despite their indirect nature, the magnitude of the band gaps remains favourable for single-junction photovoltaics (Table \ref{tab:optoelectronic}).
The band gaps follow a clear chemical trend: \ce{HZHfS3} (\SI{2.11}{eV}) $>$ \ce{HZZrS3} (\SI{1.85}{eV}) $>$ \ce{HZHfSe3} (\SI{1.63}{eV}) $>$ \ce{HZZrSe3} (\SI{1.31}{eV}).
The substitution of S for Se systematically narrows the band gap, driven by the higher energy (lower ionisation potential) of the more diffuse Se $4p$ orbitals which raises the VBM.
Conversely, substituting Zr ($4d$) with Hf ($5d$) widens the gap by approximately \SI{0.2}{\eV}, consistent with the higher orbital energy of the Hf $5d$ orbitals which raises the CBM.

\begin{table}[t]
    \centering
    \caption{Optoelectronic properties of hydrazinium-based compounds. The direct ($E_g^d$) and indirect ($E_g^i$) band gaps are given in eV, the effective masses ($m_e$ and $m_h$) are given in units of the electron rest mass. We include the high-frequency ($\varepsilon_{\infty}$) and static dielectric constants ($\varepsilon_{0}$) along the $a$, $b$, and $c$ lattice vectors. The corresponding theoretical efficiency at \SI{200}{\nm} ($\eta_{200}$) is also provided. The ionisation potential (IP) and electron affinity (EA) are given in eV.}
    \begin{adjustbox}{width=\textwidth}
    \begin{tabular}{lccccccccccccc}
    \toprule
     & $E_g^d$ & $E_g^i$ & $m_e$  & $m_h$ & \multicolumn{3}{c}{$\varepsilon_{\infty}$} & \multicolumn{3}{c}{$\varepsilon_{0}$} & $\eta_{200}$ & EA & IP \\
    \cmidrule(lr){6-8} \cmidrule(lr){9-11}
    Compound & (eV) &  (eV) & ($m_0$) & ($m_0$) & $a$ & $b$ & $c$ & $a$ & $b$ & $c$ & ($\%$) & (eV) & (eV) \\
    \midrule
    \ce{HZZrS3}  & 1.85 & 1.73 & 0.58 & 1.59 & 6.6 & 6.8 & 6.9 & 80.8 & 71.3  & 88.2   & 16.9 & 4.34 & 6.07 \\
    \ce{HZZrSe3} & 1.36 & 1.31 & 0.56 & 1.17 & 8.2 & 8.6 & 7.8 & 82.8 & 110.8 & 108.7  & 24.5 & 4.53 & 5.84 \\
    \ce{HZHfS3}  & 2.11 & 1.96 & 0.55 & 1.52 & 6.2 & 6.4 & 6.4 & 50.2 & 51.5  & 52.3   & 12.4 & 4.14 & 6.11 \\
    \ce{HZHfSe3} & 1.77 & 1.77 & 0.33 & 1.31 & 7.5 & 7.8 & 7.3 & 64.4 & 69.9  & 73.4   & 20.0 & 4.37 & 5.89 \\
    \bottomrule
    \end{tabular}
    \end{adjustbox}
    \label{tab:optoelectronic}
\end{table}

To provide an indication of carrier transport under realistic operating conditions, we calculated conductivity effective masses using the \textsc{AMSET} package.\cite{ganose_efficient_2021}
Unlike simple parabolic band fitting, this approach accounts for band non-parabolicity and multi-valley effects by computing a weighted average of the group velocities at a carrier concentration of \SI{e18}{\cm^{-3}} and \SI{300}{\kelvin}.
This formalism is essential for the hybrid chalcogenides since the symmetry-breaking octahedral tilting introduces a flat band near the conduction band minimum, a new feature that is absent in the inorganic \ce{BaZrS3}.
The presence of this heavy band, lying nearly degenerate with the conduction band minimum, increases the electron effective masses of the hybrid series (\SIrange{0.51}{0.58}{\electronmass}) relative to inorganic \ce{BaZrS3} (\SI{0.33}{\electronmass}).
For holes, the effective masses range from \SIrange{1.17}{1.59}{\electronmass} (Table \ref{tab:optoelectronic}).
Notably, the selenides exhibit lighter holes than both their sulphide counterparts and the inorganic \ce{BaZrS3} reference (\SI{1.31}{\electronmass}), driven by the increased bandwidth afforded by the Se $4p$ orbitals.
Overall, these results indicate that while the hybrid A-site induces charge localisation, the resulting transport properties remain sufficient for efficient carrier collection and reduced recombination losses in thin-film device architectures.
In particular, \ce{HZZrSe3} stands out with an ideal quasi-direct band gap value of \SI{1.31}{eV} and reasonable effective mass for both electron and holes.

\subsection{Optical Properties and Predicted Photovoltaic Performance}

\begin{figure}[t]
    \centering
    \includegraphics[width=\textwidth]{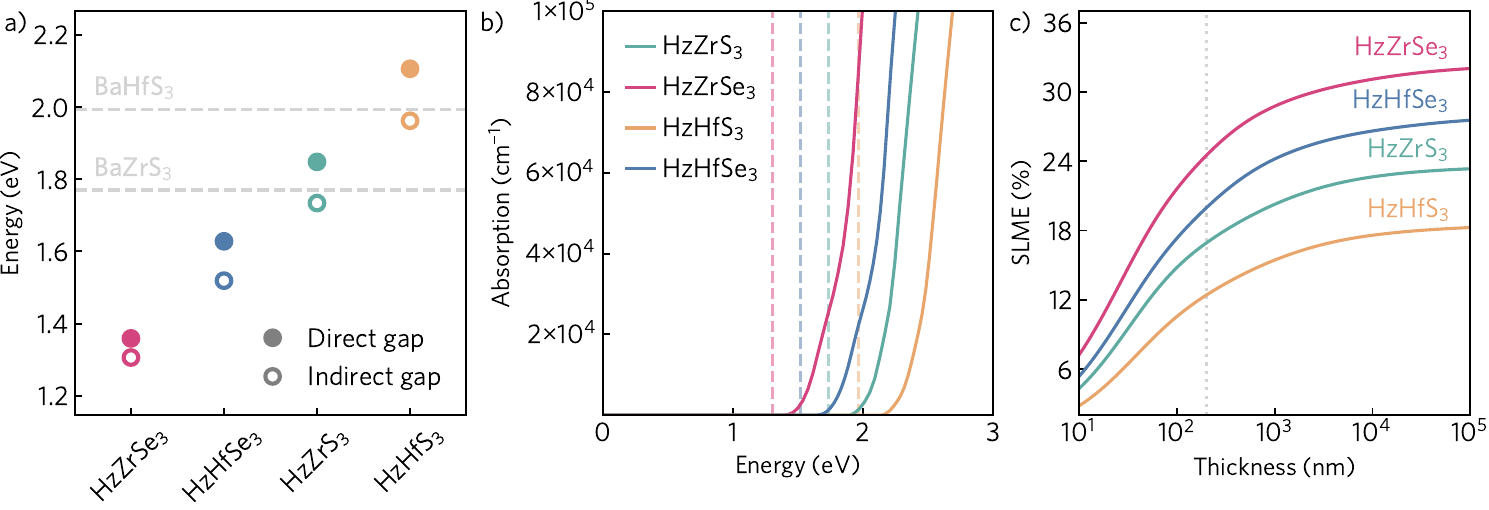}
    \caption{Optoelectronic properties of hybrid chalcogenide perovskites calculated using HSE06.
    (a) Direct (filled circles) and indirect (open circles) band gaps of the series.
    (b) Optical absorption spectra compounds as a function of energy. The dashed coloured lines represent the indirect band gap energy.
    (c) Spectroscopic Theoretical Maximum Efficiency (SLME) metric for as a function of film thickness. The dashed grey line represents a thickness of \SI{200}{\nm} as a guide to the eye.}
    \label{fig:optics}
\end{figure}

While favourable band gaps and carrier effective masses are a prerequisite for candidate absorbers, high photovoltaic performance ultimately depends on the material's ability to absorb light and generate voltage.
Figure \ref{fig:optics}b presents the optical absorption spectra for the series calculated using the HSE06 hybrid functional.
Despite the indirect nature of the fundamental band gap, the absorption onsets are strong, exceeding \SI{e5}{\per\cm} within \SI{0.5}{\eV} of the indirect gap.
This steep rise is a direct consequence of the small energy difference between the fundamental direct/indirect transitions and the presence of localised states at the valence band edge.
While weak phonon-assisted absorption is not explicitly modelled in our simulations, this will be almost immediately overtaken by strong, dipole-allowed direct transitions.
This behaviour contrasts favourably with silicon, where the large difference between indirect and direct gaps necessitates thick absorber layers to ensure sufficient absorption.

To quantify the theoretical photovoltaic performance limit, we calculated the Spectroscopic Limited Maximum Efficiency (SLME) as a function of film thickness (Figure \ref{fig:optics}c).\cite{zunger_slme_2012}
This metric accounts for the absorption of the material and the thermodynamic losses associated with an indirect band gap.
We note that non-radiative recombination is not treated explicitly and therefore this metric represents a maximum theoretical performance in the absence of other efficiency loss mechanisms.
Consistent with their band gaps, the selenide compounds exhibit higher potential efficiencies than the sulphides.
\ce{HZZrSe3} yields the highest theoretical efficiency of \SI{24.5}{\percent} for a \SI{200}{\nm} film, followed by \ce{HZHfSe3} at \SI{20.0}{\percent}.
We note that our calculated efficiency for \ce{HZZrS3} (\SI{16.9}{\percent}) is lower than the \SI{29}{\percent} reported previously by \citet{sun_chalcogenide_2015}.
This difference is attributed to the omission of the indirect band gap in their calculations, which leads to a higher overall performance prediction, along with reporting the maximum efficiency at very large thicknesses.
Indeed, we calculate a maximum theoretical efficiency of \SI{29.5}{\percent} when only the direct band gap is considered.
The predicted efficiency of \ce{HZZrSe3} significantly outperforms other emerging antimony chalcogenide absorbers such as \ce{Sb2Se3} ($\sim$\SI{10}{\percent}) \cite{wang_lone_2022a} at comparable thicknesses.

Finally, we consider the dielectric response, which plays a crucial role in charge carrier screening and defect tolerance.
Table \ref{tab:optoelectronic} summarises the high-frequency ($\varepsilon_{\infty}$) and static ($\varepsilon_{0}$) dielectric constants.
The introduction of the hydrazinium cation and the associated lattice distortion leads to significant anisotropy.
The high-frequency dielectric constants range from \numrange{6.2}{8.6}, comparable to established thin-film absorbers like CdTe ($\varepsilon_{\infty} \approx 7.1$) \cite{madelung_semiconductors_2004}.
The static dielectric constants are considerably larger, ranging from \numrange{50}{110}.
These values are substantially higher than those of typical inorganic chalcogenides and suggest effective dielectric screening of charged defects and impurities.\cite{ganose_defect_2022}
Such strong screening typically suppresses carrier capture rates and reduces the binding energy of excitons, facilitating efficient charge separation and transport even in the presence of defects.

\subsection{Band Alignments}

\begin{figure}[t]
    \centering
    \includegraphics[width=0.9\textwidth]{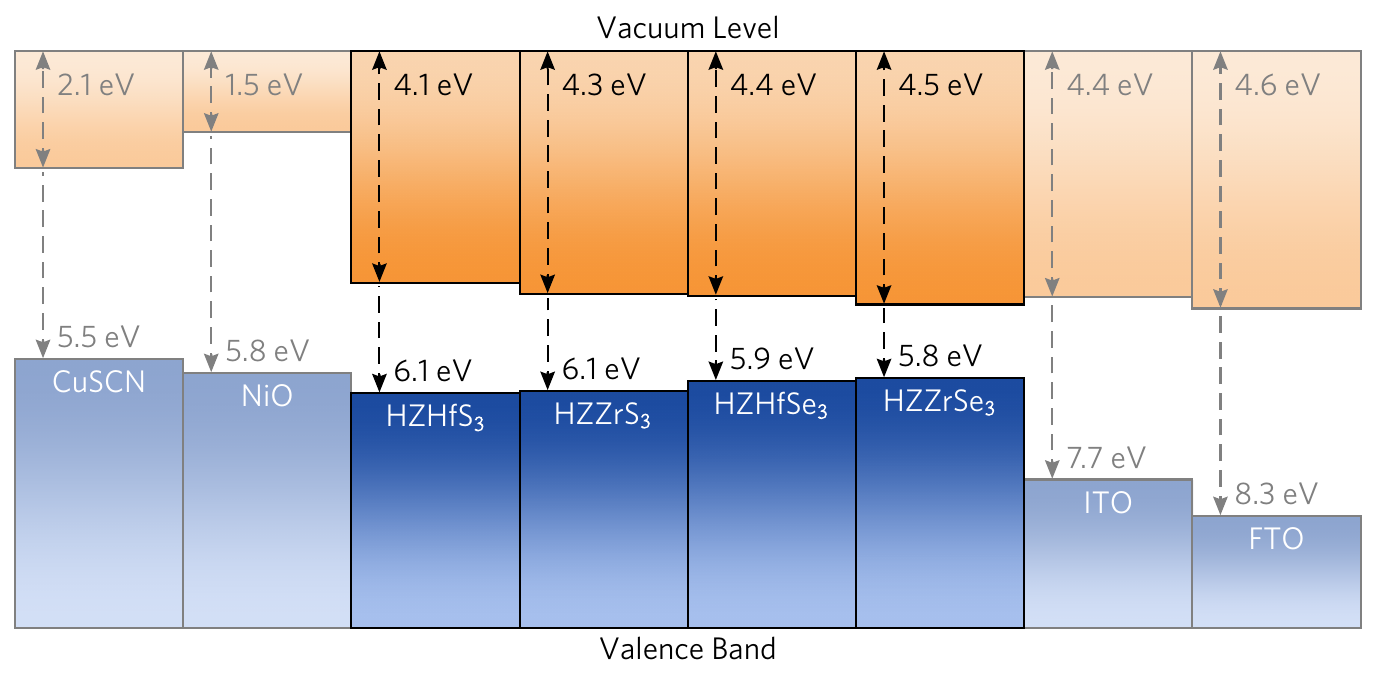}
    \caption{Comparison of absolute band alignments for hybrid chalcogenide perovskites against other well-known and emerging photovoltaic absorbers.}
    \label{fig:band_alignment}
\end{figure}

Finally, to inform the design of suitable device architectures, we calculated the band positions relative to the vacuum level.
Figure \ref{fig:band_alignment} summarises the ionisation potentials (IP) and electron affinities (EA) for the HZ series.
We find that the band positions are relatively deep compared to established absorbers, a consequence of the high electronegativity of S and Se.
The ionisation potentials range from \SI{5.8}{\eV} to \SI{6.1}{\eV}.
We observe a clear chemical trend where the sulphide compounds exhibit deeper valence bands compared to their selenide counterparts.
This \SI{\sim 0.3}{\eV} shift is driven by the lower binding energy of the Se $4p$ orbitals relative to the S $3p$ orbitals which pushes up the valence band maximum.
This trend is beneficial for device integration, as the shallower IP of the selenides facilitates hole extraction.
Conversely, the conduction band positions are governed by the B-site cation.
Hf-based compounds exhibit slightly shallower electron affinities (\ce{HZHfS3}: \SI{4.1}{\eV}) than their Zr-based analogues (\ce{HZZrS3}: \SI{4.3}{\eV}), consistent with the higher orbital energy of the Hf $5d$ orbitals.
The values are broadly comparable to the inorganic chalcogenide perovskites, for example \ce{BaZrS3} (IP = \SI{6.3}{\eV}; EA = \SI{4.6}{\eV}),\cite{nelson_bazrs3_2023} and the mixed anion absorber BiSI (IP = \SI{6.4}{\eV}; EA= \SI{4.9}{\eV}).\cite{ganose_relativistic_2016}

The electron affinities (\SIrange{4.1}{4.5}{\eV}) are well-matched to common \textit{n}-type metal oxide transport layers.
For instance, \ce{SnO2} (EA $\approx$ \SI{4.5}{\eV})\cite{Patel2025} and indium tin oxide (EA $\approx$ \SI{4.4}{\eV})\cite{Singh1978,Sugiyama2000} are likely to form favourable ohmic contacts for electron extraction, particularly for the Zr-based compounds.
However, hole extraction poses a more significant challenge.
With ionisation potentials exceeding \SI{6.0}{\eV}, common hole transport materials such as Spiro-OMeTAD or P3HT (IP $\approx$ \SI{5.0}{\eV})\cite{Zhai2025_Spiro,Ansari2018_P3HT} would create large interfacial barriers, blocking hole transport and limiting the maximum open circuit voltage.
Instead, high-work-function inorganic hole transporters will be required.
Likely candidates include \ce{NiO_x} (IP $\approx$ \SIrange{5.4}{6.0}{\eV})\cite{Wu1997} or \ce{CuSCN} (IP $\approx$ \SI{5.5}{\eV}),\cite{Kim2016_CuSCN} although interface engineering may still be needed to minimise voltage losses.

\section{Conclusion}

In this work, we have employed first-principles calculations to systematically assess the viability of hybrid chalcogenide perovskites as stable, lead-free alternatives for thin-film photovoltaics.
Through a hierarchical screening of 21 organic A-site cations, we identified the hydrazinium family, HZ(Zr,Hf)(S,Se)$_3$, as the only candidates capable of supporting a stable perovskite framework.
Thermodynamic analysis and molecular dynamics simulations suggest that these compounds are stable against decomposition and maintain dynamical stability at room temperature.
Furthermore, we demonstrate that while these systems possess indirect fundamental band gaps, the degree of indirectness is small (\SIrange{0.05}{0.15}{\eV}), rendering these materials effectively quasi-direct absorbers at room temperature.
This is reflected in their optical properties, which exhibit absorption coefficients exceeding \SI{e5}{\per\cm} just above the band edge and yield theoretical efficiencies up to \SI{24.5}{\percent} for \SI{200}{\nm} \ce{HZZrSe3} films.
Furthermore, despite the presence of a flat band introduced by octahedral tilting, the carrier effective masses remain sufficiently low to support efficient charge transport.

The primary challenge for hybrid chalcogenide perovskites lies in their chemical constraints and device integration.
The strict steric requirement for small A-site cations ($R_A < \SI{217}{pm}$) limits the chemical space available for tuning, while the deep ionisation potentials ($\sim$\SI{6}{\eV}) necessitate the development of high-work-function hole contact materials.
Furthermore, since these are purely hypothetical compounds, a major challenge will be identifying feasible synthesis routes.
We propose a solution-based route utilising hydrazinium sulphate (\ce{N2H6SO4}) as a precursor, which can be prepared by reacting hydrazinium azide (\ce{[N2H5][N3]}) with sulphuric acid (\ce{H2SO4}) in an aqueous solution, as outlined by \citet{klapotke_reaction_1996}:
\begin{equation}
    \ce{[N2H5][N3] + H2SO4 -> N2H6SO4 + NH3}
\end{equation}
This salt can be used in a solid-state reaction with zirconium disulphide (\ce{ZrS2}) and hydrogen sulphide (\ce{H2S}) to synthesise \ce{HZZrS3}, analogous to existing chalcogenide synthesis methods:\cite{tiwari_chalcogenide_2021, wang_synthesis_2001,hahn_untersuchungen_1957}
\begin{equation}
    \ce{N2H6SO4 + ZrS2 + H2S -> N2H6ZrS3 + H2SO4}
\end{equation}
If these synthetic and interfacial challenges can be addressed, hybrid hydrazinium cations could offer a new direction for increasing the tunability of chalcogenide perovskite photovoltaics.

\section{Acknowledgements}

A.M.G.~was supported by EPSRC Fellowship EP/T033231/1. We are grateful to the UK Materials and Molecular Modelling Hub for computational resources, which are partially funded by EPSRC (EP/T022213/1, EP/W032260/1 and EP/P020194/1).
We acknowledge computational resources and support provided by the Imperial College Research Computing Service (\url{http://doi.org/10.14469/hpc/2232}).

\bibliography{refs}

\end{document}


\section{Inorganic Chalcogenide Perovskites}
\label{sec:Inorganic}

This section presents the results for two inorganic chalcogenide perovskites, \ce{BaZrS3} and \ce{BaHfS3}, selected due to their prior experimental validation.\cite{gupta_environmentally_2020, wei_realization_2020, mitzi_high_2019, nishigaki_extraordinary_2020, hanzawa_material_2019} Comparing the computational results in this study with established experimental data serves to validate the accuracy of our calculation parameters. Once verified, these parameters can be applied to hybrid compounds to predict the properties of novel chalcogenide perovskites lacking experimental confirmation, ensuring the reliability and precision of the computational methods employed.
Table \ref{stab:inorganic_comp_params} shows the converged properties for the inorganic chalcogenides mentioned (\ce{BaZrS3} and \ce{BaHfS3}), adhering to the geometric and electronic parameters described in the main text.

\begin{table}[h]
\centering
\caption{Calculation parameters for \ce{BaZrS3} and \ce{BaHfS3}.}
\begin{tabular}{c c c c c}
\toprule
Compound & Structure & Space group & Cut-off (eV) & $k$-point mesh \\
\midrule
\ce{BaZrS3} & Orthorhombic & \textit{Pnma} & 300 & $3\times3\times4$ \\
\ce{BaHfS3} & Orthorhombic & \textit{Pnma} & 300 & $4\times3\times4$ \\
\bottomrule
\label{stab:inorganic_comp_params}
\end{tabular}
\end{table}

\begin{figure}[h]
    \centering
    \includegraphics[width=0.8\textwidth]{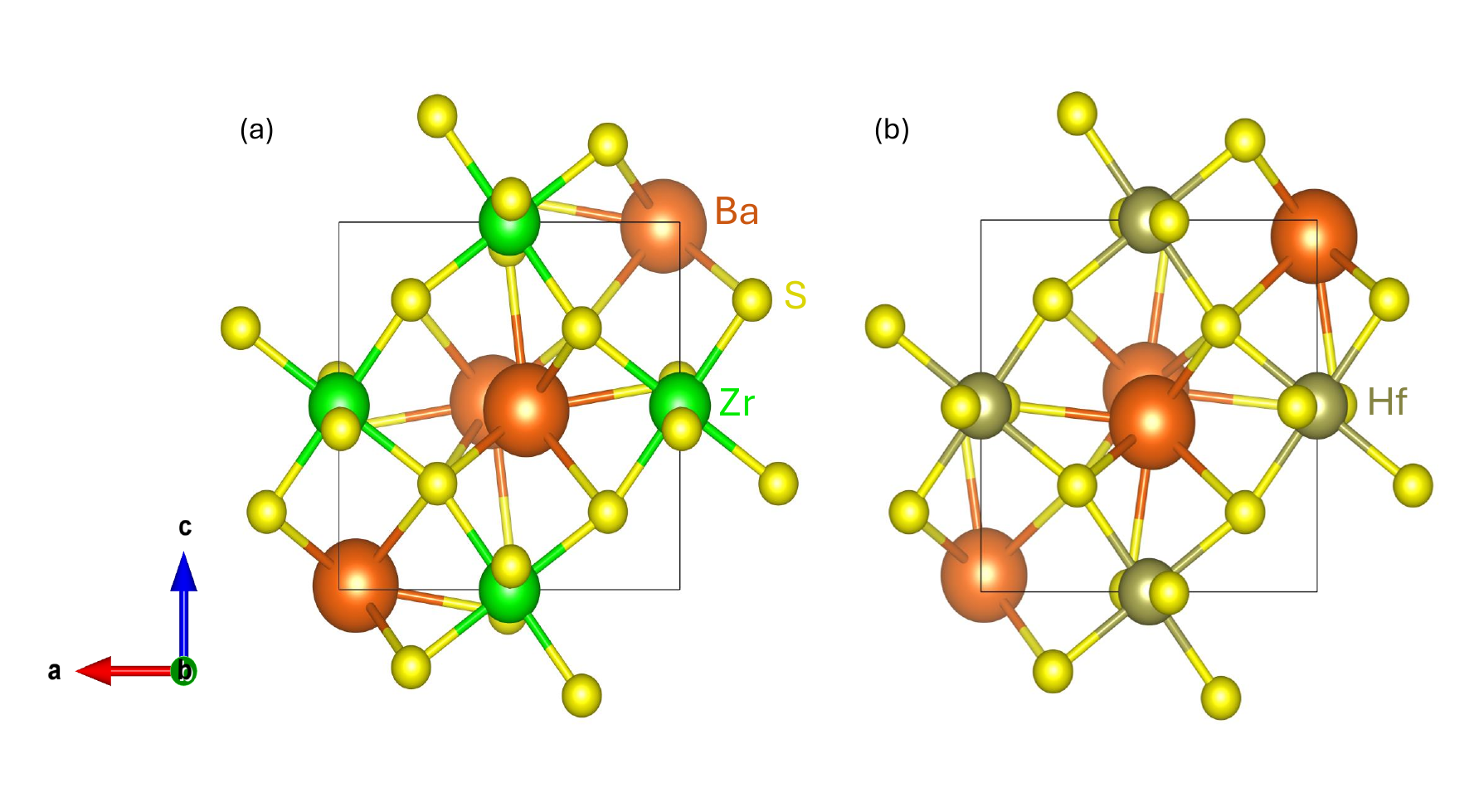}
    \caption{Optimised structure of (a) \ce{BaZrS3} and (b) \ce{BaHfS3}.}
    \label{sfig:inorganic_crystal_structure}
\end{figure}

The geometries of these structures were optimised using the PBEsol functional. The associated lattice parameters are listed in Table \ref{stab:inorganic_latt_param}. In both compounds, the calculated lattice parameters are within \SI{1}{\percent} of experimental values, confirming the parameters are suitable for subsequent use on theoretical compounds. Figure \ref{sfig:inorganic_crystal_structure} shows the optimised structures of \ce{BaZrS3} and \ce{BaHfS3}. Both chalcogenide perovskites exhibit $a^-b^+a^-$ octahedral tilting in the orthorhombic phase, reducing symmetry and altering bond angles and lengths.

\begin{table}[h]
\centering
\caption{Calculated lattice parameters and respective deviations from experimental values.\cite{nishigaki_extraordinary_2020}}
\begin{tabular}{cccc}
\toprule
Compound & $a$ (\si{\angstrom}) & $b$ (\si{\angstrom}) & $c$ (\si{\angstrom}) \\
\midrule
\ce{BaZrS3} & 7.074 ($+0.18\%$) & 9.923 ($-0.56\%$) & 6.962 ($-0.88\%$) \\
\ce{BaHfS3} & 6.998 ($-0.15\%$) & 9.862 ($-0.56\%$) & 6.938 ($-0.91\%$) \\
\bottomrule
\label{stab:inorganic_latt_param}
\end{tabular}
\end{table}

\begin{figure}[h]
    \centering
    \includegraphics[width=0.8\textwidth]{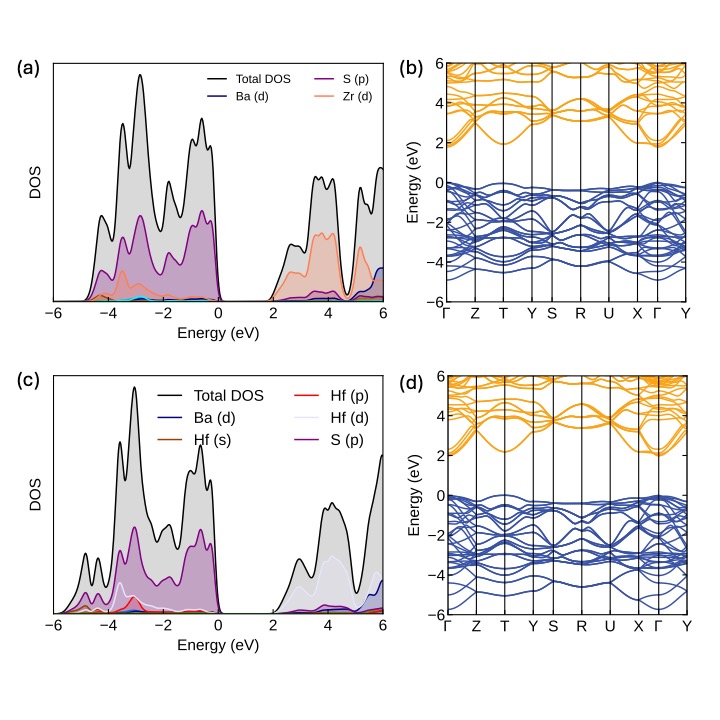}
    \caption{Partial density of states and band structure of (a--b) \ce{BaZrS3} and (c--d) \ce{BaHfS3}.}
    \label{sfig:inorganic_electronic}
\end{figure}

The density of states (DOS) and band structures were calculated using the HSE06 functional for both \ce{BaZrS3} and \ce{BaHfS3} and are presented in Figure \ref{sfig:inorganic_electronic}. The DOS of \ce{BaZrS3} indicates a band gap of \SI{1.77}{eV}, aligning closely with experimental results of \SI{1.75}{eV}\cite{gupta_environmentally_2020} and \SI{1.79}{eV},\cite{osei-agyemang_examining_2021} confirming the accuracy of the HSE06 functional.

The band structure in Figure \ref{sfig:inorganic_electronic}b depicts the valence band in blue and the conduction band in orange. The regions of interest within the band structures are around the band edges, which dictate the band gap and are in agreement with the DOS plot. The valence band maximum (VBM) occurs at the $\Gamma$-point (0, 0, 0) in $k$-space. The occupied valence bands from \SI{0}{eV} to \SI{-2}{eV} in \ce{BaZrS3} lack dispersion as they are relatively flat, due to the localised S $p$ states. The conduction band minimum (CBM) also lies on the $\Gamma$-point, identifying \ce{BaZrS3} as a direct band gap semiconductor. The CBM displays more dispersive bands at energy levels from \SIrange{1.78}{3}{eV}. This can be attributed to the broader DOS resulting from an overlap of Zr $d$-states with S $p$-states, leading to delocalisation and higher dispersion. The higher dispersion results in a lower electron effective mass ($m_e$) of \SI{0.38}{\electronmass} compared to a hole effective mass ($m_h$) of \SI{0.48}{\electronmass} in the less dispersed valence band, aligning well with the results of \citet{huo_high-throughput_2018}.

Figure \ref{sfig:inorganic_electronic}c shows the DOS plot for \ce{BaHfS3}, where the band gap is slightly larger with a value of \SI{1.98}{eV}, agreeing well with experimental band gaps of \SI{2.06}{eV}.\cite{hanzawa_material_2019} The band structure in Figure \ref{sfig:inorganic_electronic}d is similar to that of \ce{BaZrS3}, with strong dispersions in the CBM resulting in an electron effective mass of \SI{0.29}{\electronmass}. The VB, on the other hand, exhibits a hole effective mass of \SI{2.41}{\electronmass}. This increase in $m_h$ can be attributed to the VBM being located at the T-point rather than the $\Gamma$-point, making \ce{BaHfS3} a slightly indirect semiconductor; however, it retains desirable absorption properties due to the presence of a direct transition at \SI{1.99}{eV} at the $\Gamma$-point. Table \ref{stab:inorganic_electronic} summarises the electronic properties of Ba(Zr,Hf)S$_3$.

\begin{table}[h]
    \centering
    \caption{Optoelectronic properties of inorganic chalcogenide perovskites. The direct ($E_g^d$) and indirect ($E_g^i$) band gaps are given in eV, the effective masses ($m_e$ and $m_h$) are given in units of the electron rest mass. We include the high-frequency dielectric constant ($\varepsilon_{\infty}$) along the $a$, $b$, and $c$ lattice vectors. The corresponding theoretical efficiency at \SI{200}{\nm} ($\eta_{200}$) is also provided.}
    \begin{tabular}{lcccccccccc}
    \toprule
     & $E_g^d$ & $E_g^i$ & $m_e$  & $m_h$ & \multicolumn{3}{c}{$\varepsilon_{\infty}$} & $\eta_{200}$ \\
    \cmidrule(lr){6-8} 
    Compound & (eV) &  (eV) & ($m_0$) & ($m_0$) & $a$ & $b$ & $c$ & ($\%$) \\
    \midrule
    \ce{BaZrS3} & 1.77 & -- & 0.30 & 0.48 & 7.27 & 7.31 & 7.13 & 26.81 \\
    \ce{BaHfS3} & 1.99 & 1.98 & 0.29 & 2.41  & 6.57 & 6.67 & 6.60 & 23.0 \\
    \bottomrule
    \label{stab:inorganic_electronic}
    \end{tabular}
\end{table}

\begin{figure}[h]
    \centering
    \includegraphics[width=0.4\textwidth]{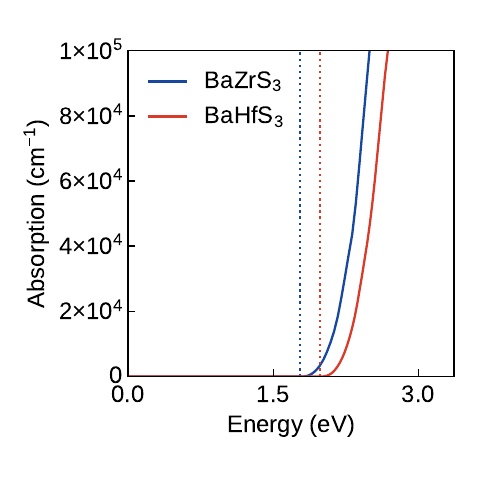}
    \caption{HSE06 calculated optical absorption spectra for \ce{BaZrS3} and \ce{BaHfS3} as a function of energy. Solid lines indicate the optical absorption and dashed lines indicate the fundamental band gap.}
    \label{sfig:inorganic_optics}
\end{figure}

\begin{figure}[h]
    \centering
    \includegraphics[width=0.4\textwidth]{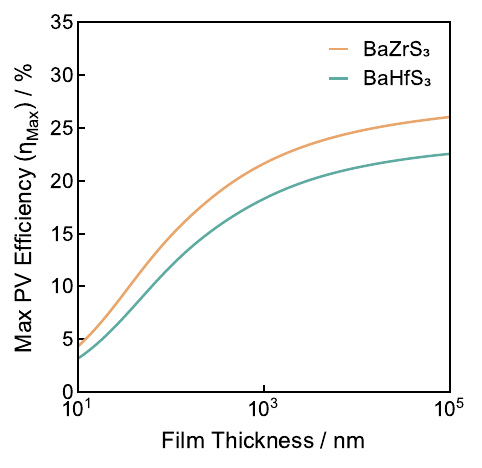}
    \caption{Maximum theoretical efficiency for \ce{BaZrS3} and \ce{BaHfS3} as a function of film thickness, calculated using HSE06.}
    \label{sfig:inorganic_slme}
\end{figure}

The absorption spectra for Ba(Zr,Hf)S$_3$ are shown in Figure \ref{sfig:inorganic_optics}. Both compounds exhibit strong absorption coefficients, represented by a sharp absorption edge which climbs to above \SI{e5}{\per\cm} within \SI{0.5}{eV} of the onset, agreeing well with experimental literature.\cite{nishigaki_extraordinary_2020, wei_realization_2020, marquez_bazrs3_2021, qin_promising_2023} Additionally, the high-frequency dielectric constants, $\varepsilon_{\infty}$, and maximum theoretical efficiency were calculated and are listed in Table \ref{stab:inorganic_electronic} and Figure \ref{sfig:inorganic_slme}.

\newpage
\section{Materials Screening}

{
\singlespacing
\begin{longtable}{lrrrrr}
\caption{Tolerance factors ($t$) and octahedral factors ($\mu$) for hybrid chalcogenide perovskites. Compounds that remain perovskite after relaxation are highlighted in bold.} \label{stab:tolerance_factors} \\

\toprule
Compound & $R_A$ (pm) & $R_B$ (pm) & $R_X$ (pm) & $t$ & $\mu$ \\
\midrule
\endfirsthead

\multicolumn{6}{c}{{\bfseries \tablename\ \thetable{} -- continued from previous page}} \\
\toprule
Compound & $R_A$ (pm) & $R_B$ (pm) & $R_X$ (pm) & $t$ & $\mu$ \\
\midrule
\endhead

\midrule
\multicolumn{6}{r}{{Continued on next page...}} \\
\endfoot

\bottomrule
\endlastfoot

\textbf{HZZrS$_3$} & 217 & 72 & 184 & 1.11 & 0.39 \\
AMZZrS$_3$ & 267 & 72 & 184 & 1.25 & 0.39 \\
\textbf{HZZrSe$_3$} & 217 & 72 & 196 & 1.09 & 0.37 \\
AMZZrSe$_3$ & 267 & 72 & 196 & 1.22 & 0.37 \\
\textbf{HZHfS$_3$} & 217 & 71 & 184 & 1.11 & 0.39 \\
AMZHfS$_3$ & 267 & 71 & 184 & 1.25 & 0.39 \\
\textbf{HZHfSe$_3$} & 217 & 71 & 196 & 1.09 & 0.36 \\
AMZHfSe$_3$ & 267 & 71 & 196 & 1.23 & 0.36 \\
AMSbS$_3$ & 146 & 60 & 184 & 0.96 & 0.33 \\
HYSbS$_3$ & 140 & 60 & 184 & 0.94 & 0.33 \\
MAMSbS$_3$ & 216 & 60 & 184 & 1.16 & 0.33 \\
MASbS$_3$ & 217 & 60 & 184 & 1.16 & 0.33 \\
HZSbS$_3$ & 220 & 60 & 184 & 1.17 & 0.33 \\
HASbS$_3$ & 216 & 60 & 184 & 1.16 & 0.33 \\
PHSbS$_3$ & 212 & 60 & 184 & 1.15 & 0.33 \\
AZSbS$_3$ & 238 & 60 & 184 & 1.22 & 0.33 \\
FASbS$_3$ & 253 & 60 & 184 & 1.27 & 0.33 \\
EASbS$_3$ & 274 & 60 & 184 & 1.33 & 0.33 \\
FOSbS$_3$ & 258 & 60 & 184 & 1.28 & 0.33 \\
DMASbS$_3$ & 272 & 60 & 184 & 1.32 & 0.33 \\
MHSbS$_3$ & 275 & 60 & 184 & 1.33 & 0.33 \\
AZESbS$_3$ & 250 & 60 & 184 & 1.26 & 0.33 \\
HMASbS$_3$ & 275 & 60 & 184 & 1.33 & 0.33 \\
CASbS$_3$ & 293 & 60 & 184 & 1.38 & 0.33 \\
TMASbS$_3$ & 294 & 60 & 184 & 1.39 & 0.33 \\
PYSbS$_3$ & 278 & 60 & 184 & 1.34 & 0.33 \\
TTMASbS$_3$ & 292 & 60 & 184 & 1.38 & 0.33 \\
AMSbSe$_3$ & 146 & 60 & 196 & 0.94 & 0.31 \\
HYSbSe$_3$ & 140 & 60 & 196 & 0.93 & 0.31 \\
MAMSbSe$_3$ & 216 & 60 & 196 & 1.14 & 0.31 \\
MASbSe$_3$ & 217 & 60 & 196 & 1.14 & 0.31 \\
HZSbSe$_3$ & 220 & 60 & 196 & 1.15 & 0.31 \\
HASbSe$_3$ & 216 & 60 & 196 & 1.14 & 0.31 \\
PHSbSe$_3$ & 212 & 60 & 196 & 1.13 & 0.31 \\
AZSbSe$_3$ & 238 & 60 & 196 & 1.20 & 0.31 \\
FASbSe$_3$ & 253 & 60 & 196 & 1.24 & 0.31 \\
EASbSe$_3$ & 274 & 60 & 196 & 1.30 & 0.31 \\
FOSbSe$_3$ & 258 & 60 & 196 & 1.25 & 0.31 \\
DMASbSe$_3$ & 272 & 60 & 196 & 1.29 & 0.31 \\
MHSbSe$_3$ & 275 & 60 & 196 & 1.30 & 0.31 \\
AZESbSe$_3$ & 250 & 60 & 196 & 1.23 & 0.31 \\
HMASbSe$_3$ & 275 & 60 & 196 & 1.30 & 0.31 \\
CASbSe$_3$ & 293 & 60 & 196 & 1.35 & 0.31 \\
TMASbSe$_3$ & 294 & 60 & 196 & 1.35 & 0.31 \\
PYSbSe$_3$ & 278 & 60 & 196 & 1.31 & 0.31 \\
TTMASbSe$_3$ & 292 & 60 & 196 & 1.35 & 0.31 \\
AMBiS$_3$ & 146 & 76 & 184 & 0.90 & 0.41 \\
\textbf{HYBiS$_3$} & 140 & 76 & 184 & 0.88 & 0.41 \\
MAMBiS$_3$ & 216 & 76 & 184 & 1.09 & 0.41 \\
MABiS$_3$ & 217 & 76 & 184 & 1.09 & 0.41 \\
HZBiS$_3$ & 220 & 76 & 184 & 1.10 & 0.41 \\
HABiS$_3$ & 216 & 76 & 184 & 1.09 & 0.41 \\
PHBiS$_3$ & 212 & 76 & 184 & 1.08 & 0.41 \\
AZBiS$_3$ & 238 & 76 & 184 & 1.15 & 0.41 \\
FABiS$_3$ & 253 & 76 & 184 & 1.19 & 0.41 \\
EABiS$_3$ & 274 & 76 & 184 & 1.25 & 0.41 \\
FOBiS$_3$ & 258 & 76 & 184 & 1.20 & 0.41 \\
DMABiS$_3$ & 272 & 76 & 184 & 1.24 & 0.41 \\
MHBiS$_3$ & 275 & 76 & 184 & 1.25 & 0.41 \\
AZEBiS$_3$ & 250 & 76 & 184 & 1.18 & 0.41 \\
HMABiS$_3$ & 275 & 76 & 184 & 1.25 & 0.41 \\
CABiS$_3$ & 293 & 76 & 184 & 1.30 & 0.41 \\
TMABiS$_3$ & 294 & 76 & 184 & 1.30 & 0.41 \\
PYBiS$_3$ & 278 & 76 & 184 & 1.26 & 0.41 \\
TTMABiS$_3$ & 292 & 76 & 184 & 1.29 & 0.41 \\
AMBiSe$_3$ & 146 & 76 & 196 & 0.89 & 0.39 \\
HYBiSe$_3$ & 140 & 76 & 196 & 0.87 & 0.39 \\
MAMBiSe$_3$ & 216 & 76 & 196 & 1.07 & 0.39 \\
MABiSe$_3$ & 217 & 76 & 196 & 1.07 & 0.39 \\
HZBiSe$_3$ & 220 & 76 & 196 & 1.08 & 0.39 \\
HABiSe$_3$ & 216 & 76 & 196 & 1.07 & 0.39 \\
PHBiSe$_3$ & 212 & 76 & 196 & 1.06 & 0.39 \\
AZBiSe$_3$ & 238 & 76 & 196 & 1.13 & 0.39 \\
FABiSe$_3$ & 253 & 76 & 196 & 1.17 & 0.39 \\
EABiSe$_3$ & 274 & 76 & 196 & 1.22 & 0.39 \\
FOBiSe$_3$ & 258 & 76 & 196 & 1.18 & 0.39 \\
DMABiSe$_3$ & 272 & 76 & 196 & 1.22 & 0.39 \\
MHBiSe$_3$ & 275 & 76 & 196 & 1.22 & 0.39 \\
AZEBiSe$_3$ & 250 & 76 & 196 & 1.16 & 0.39 \\
HMABiSe$_3$ & 275 & 76 & 196 & 1.22 & 0.39 \\
CABiSe$_3$ & 293 & 76 & 196 & 1.27 & 0.39 \\
TMABiSe$_3$ & 294 & 76 & 196 & 1.27 & 0.39 \\
PYBiSe$_3$ & 278 & 76 & 196 & 1.23 & 0.39 \\
TTMABiSe$_3$ & 292 & 76 & 196 & 1.27 & 0.39 \\
\end{longtable}
}

\begin{table}[h]
\centering
\caption{The 46 compounds examined in this report with their converged $k$-points and phases formed after geometric optimisation and electronic convergence for the initial screening. Note, for the stable perovskites, the $k$-point mesh was reconverged following ground state structure determination from molecular dynamics annealing. The cut-off energy for all compounds was \SI{500}{eV}.}
\begin{tabular}{lcc|lcc}
\toprule
\multicolumn{3}{c}{\ce{ABS3} perovskites} & \multicolumn{3}{c}{\ce{ABSe3} perovskites} \\
\midrule
Compound & $k$-point & Phase & Compound & $k$-point & Phase \\
\midrule
\ce{HYBiS3} & $8\times 8\times 8$ & Perovskite & \ce{HYBiSe3} & $8\times 8\times 8$ & Non-perovskite \\
\ce{AMBiS3} & $9\times 9\times 9$ & Non-perovskite & \ce{AMBiSe3} & $8\times 8\times 8$ & Non-perovskite \\
\ce{PHBiS3} & $9\times 9\times 9$ & Non-perovskite & \ce{PHBiSe3} & $9\times 9\times 9$ & Non-perovskite \\
\ce{HABiS3} & $9\times 9\times 9$ & Non-perovskite & \ce{HABiSe3} & $6\times 6\times 6$ & Non-perovskite \\
\ce{MABiS3} & $6\times 6\times 6$ & Non-perovskite & \ce{MABiSe3} & $6\times 6\times 6$ & Non-perovskite \\
\ce{HZBiS3} & $7\times 7\times 7$ & Non-perovskite & \ce{HZBiSe3} & $7\times 7\times 7$ & Non-perovskite \\
\ce{FABiS3} & $9\times 9\times 9$ & Non-perovskite & \ce{FABiSe3} & $10\times 10\times 10$ & Non-perovskite \\
\ce{FOBiS3} & $8\times 8\times 8$ & Non-perovskite & \ce{FOBiSe3} & $7\times 7\times 7$ & Non-perovskite \\
\ce{EABiS3} & $7\times 7\times 7$ & Non-perovskite & \ce{EABiSe3} & $7\times 7\times 7$ & Non-perovskite \\
\ce{AZBiS3} & $7\times 7\times 7$ & Non-perovskite & \ce{AZBiSe3} & $7\times 7\times 7$ & Non-perovskite \\
\ce{AZEBiS3} & $8\times 8\times 8$ & Non-perovskite & \ce{AZEBiSe3} & $9\times 9\times 9$ & Non-perovskite \\
\ce{DMABiS3} & $7\times 7\times 7$ & Non-perovskite & \ce{DMABiSe3} & $7\times 7\times 7$ & Non-perovskite \\
\ce{TMABiS3} & $10\times 10\times 10$ & Non-perovskite & \ce{TMABiSe3} & $5\times 5\times 5$ & Non-perovskite \\
\ce{TTMABiS3} & $5\times 5\times 5$ & Non-perovskite & \ce{TTMABiSe3} & $5\times 5\times 5$ & Non-perovskite \\
\ce{CABiS3} & $8\times 8\times 8$ & Non-perovskite & \ce{CABiSe3} & $8\times 8\times 8$ & Non-perovskite \\
\ce{PYBiS3} & $6\times 6\times 6$ & Non-perovskite & \ce{PYBiSe3} & $7\times 7\times 7$ & Non-perovskite \\
\ce{MHBiS3} & $7\times 7\times 7$ & Non-perovskite & \ce{MHBiSe3} & $7\times 7\times 7$ & Non-perovskite \\
\ce{MAMBiS3} & $6\times 6\times 6$ & Non-perovskite & \ce{MAMBiSe3} & $9\times 9\times 9$ & Non-perovskite \\
\ce{HMABiS3} & $9\times 9\times 9$ & Non-perovskite & \ce{HMABiSe3} & $9\times 9\times 9$ & Non-perovskite \\
\midrule
\ce{HZZrS3} & $6\times 6\times 6$ & Perovskite & \ce{HZZrSe3} & $5\times 5\times 5$ & Perovskite \\
\ce{HZHfS3} & $6\times 6\times 6$ & Perovskite & \ce{HZHfSe3} & $5\times 5\times 5$ & Perovskite \\
\ce{AMZZrS3} & $5\times 5\times 5$ & Non-perovskite & \ce{AMZZrSe3} & $5\times 5\times 5$ & Non-perovskite \\
\ce{AMZHfS3} & $6\times 6\times 6$ & Non-perovskite & \ce{AMZHfSe3} & $5\times 5\times 5$ & Non-perovskite \\
\bottomrule
\end{tabular}
\label{stab:hybrid_comp_params}
\end{table}

\begin{table}
\centering
\caption{Lattice parameters for the thermodynamically and dynamically stable chalcogenide perovskites calculated using the PBEsol functional.}
\small
\begin{tabular}{ccccccc}
\toprule
Compound & $a$ (\si{\angstrom}) & $b$ (\si{\angstrom}) & $c$ (\si{\angstrom}) & $\alpha$ (\si{\degree}) & $\beta$ (\si{\degree}) & $\gamma$ (\si{\degree}) \\
\midrule
\ce{HZZrS3}  &  9.9952 & 10.1295 & 5.0957 & 90 & 90 & 90 \\
\ce{HZZrSe3} & 10.4382 & 10.4845 & 5.3321 & 90 & 90 & 90 \\
\ce{HZHfS3}  &  9.9445 & 10.0595 & 5.0581 & 90 & 90 & 90 \\
\ce{HZHfSe3} & 10.3763 & 10.4272 & 5.2987 & 90 & 90 & 90 \\
\bottomrule
\label{stab:hybrid_lattice_params}
\end{tabular}
\end{table}

\begin{figure}
    \centering
    \includegraphics[width=1\textwidth]{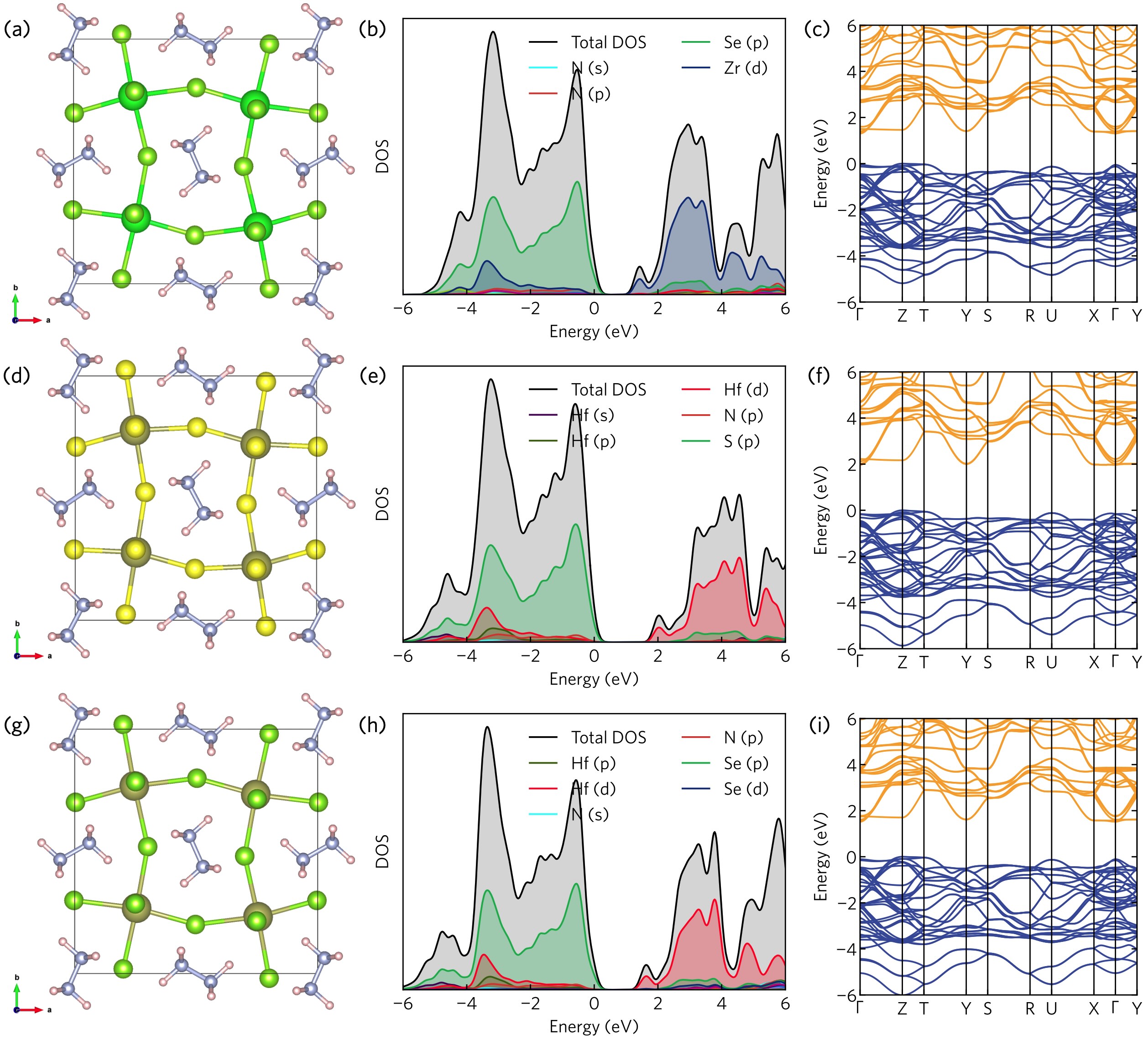}
    \caption{Atomic structures, partial density of states and band structures for (a--c) \ce{HZZrSe3}, (d--f) \ce{HZHfS3} and (g--i) \ce{HZHfSe3}.}
    \label{sfig:hybrid_electronic}
\end{figure}

\clearpage
\bibliography{refs}